\documentclass[aps,showpacs,prb,floatfix,twocolumn]{revtex4}
\usepackage{amsmath,amssymb,graphicx,bm,epsfig}

\renewcommand{\a}{\alpha}
\renewcommand{\b}{\beta}
\newcommand{\vk}{{\mathbf k}}

\begin{document}

\title{Interpolative Approach for Solving Quantum Impurity Model Based on
the Slave--Boson Mean--Field Approximation}
\author{V. Oudovenko$^{a,e}$, K. Haule$^{b,e}$, S. Y. Savrasov$^{c}$, D.
Villani$^{d,e}$, G. Kotliar$^{e}$}
\affiliation{$^{a}$Laboratory for Theoretical Physics, Joint Institute for Nuclear
Research, 141980 Dubna, Russia}
\affiliation{$^{b}$Jo\v zef Stefan Institute, SI-1000 Ljubljana, Slovenia}
\affiliation{$^{c}$Department of Physics, New Jersey Institute of Technology, Newark, NJ
07102, USA}
\affiliation{$^{d}$Hess Energy Trading Company, LLC, New York, NY 10036, USA}
\affiliation{$^{e}$Department of Physics and Center for Material Theory, Rutgers
University, Pscataway, NJ 08854, USA}
\date{\today}

\begin{abstract}
A rational representation for the self-energy is explored to interpolate the
solution of the Anderson impurity model in general orbitally degenerate
case. Several constrains such as the Friedel's sum rule, high--frequency
moments and the value of quasiparticle residue are used to establish the
equations for the coefficients of the interpolation. We test two fast
techniques, the slave--boson mean--field and the Hubbard I approximation to
determine the coefficients. The obtained self--energies are compared with
the results of numerically exact Quantum Monte Carlo method. We find that
using the slave--boson mean--field approach we can construct an accurate
self--energy for all frequencies via the proposed interpolation procedure.
\end{abstract}

\maketitle

\section{Introduction}

There has been recent progress in understanding physics of strongly
correlated electronic systems and their electronic structure near a
localization--delocalization transition through the development of dynamical
mean--field theory (DMFT) \cite{DMFT}. Merging this computationally
tractable many--body technique with realistic local--density--approximation
(LDA)\ \cite{LDA} based electronic structure calculations of strongly
correlated solids is promising due to its simplicity and correctness in both
band and atomic limits. At present, much effort is being made in this
direction including the developments of a LDA+DMFT method \cite%
{AnisimovKotliar}, LDA++ approach \cite{LDA++}, combined GW and DMFT theory 
\cite{Georges} as well as applications to various systems such as La$_{1-x}$%
Sr$_{x}$TiO$_{3}$ \cite{LaTiO3}, V$_{2}$O$_{3}$ \cite{V2O3}, Fe and Ni \cite%
{Licht}, Ce \cite{Ce}, Pu \cite{Nature,Science}, and many others. For a
review, see Ref. \onlinecite{reviews}.

The development of \textit{ab initio} DMFT scheme requires fast methods and
algorithms to solve the Anderson impurity model \cite{AIM} in general
multiorbital case. Present techniques based on either non--crossing
approximation (NCA)\ or iterative perturbation theory (IPT) are unable to
provide the solution to that problem due to a limited number of regimes
where these methods can be applied \cite{DMFT}. The Quantum Monte Carlo
(QMC) technique \cite{DMFT,Jarrell} is very accurate and can cope with
multiorbital situation but not with multiplet interactions. Also its
applicability so far has been limited either to a small number of orbitals
or to unphysically large temperatures due to its computational cost.
Recently some progress has been achieved using impurity solvers that improve
upon the NCA approximation \cite{rotors,jeschke,Haule:2001}, but it has not
been possible to retrieve Fermi liquid behavior at very low temperatures
with these methods in the orbitally degenerate case.

In this paper we explore the possibility to use interpolation for the
self--energy of the quantum impurity model in general multiorbital
situation. We do not attempt to develop an alternative method for solving
the impurity problem, but follow the ideology of LDA theory where
approximations were designed by analytical fits \cite{Vosko} to the Quantum
Monte Carlo simulations for homogeneous electron gas \cite{Ceperley}.
Numerically very expensive QMC calculations for the impurity model display
smooth self--energies at imaginary frequencies for a wide range of
interactions and dopings, and it is therefore tempting to design such an
interpolation. We also keep in mind that for many applications a high
precision in reproducing the self--energies may not be required. One of such
applications is, for example, the calculation of the total energy \cite%
{Ce,Nature,Science} which, as well known from LDA based experience, may not
be so sensitive to the details of the one--electron spectra. As a result, we
expect that even crude evaluations of the self--energy shapes on imaginary
frequency axis may be sufficient for solving many realistic total energy
problems. Another point is a computational efficiency. Bringing full
self--consistent loops with respect to charge densities \cite{Nature} and
other spectral functions require many iterations towards the convergency
which may not need too accurate frequency resolutions at every step. This
view should, of course, be contrasted with calculation of properties such as
the low--energy spectroscopy and especially transport where delicate
distribution of spectral weight at low energy, namely the imaginary part of
the analytically continued self--energy, needs to be computed with much
greater precision. Here, extensions of the interpolative algorithms should
be implemented and its beyond the scope of the present work.

We can achieve a fast interpolative algortihm for the self--energy utilizing
a rational representation. The coefficients in this interpolation can be
found by forcing the self--energy to obey several limits and constrains. For
example, if infinite frequency (Hartree--Fock) limit, high--frequency
moments of the self--energy, low--frequency mass renormalization, number of
particles as well as the value of the self--energy at zero frequency are
known from independent calculation, the set of interpolating coefficients is
well defined. In this work, we explore the slave--boson Gutzwiller approach 
\cite{Gutz,Ruck, Fleszar,Hasegawa} and the Hubbard I approximation \cite%
{Hubbard1} to determine these coefficients. Comparing interpolating results
with more accurate calculations using the Quantum Monte Carlo method we find
that the slave--boson approach predicts the parameters of interpolation with
a good accuracy while the Hubbard I method fails in a number of regimes.
Thus, interpolation for the self--energy at all frequencies can be obtained
from the method designed to work exclusively at low--frequency limit.

Our paper is organized as follows. In Section II we discuss rational
interpolation for the self--energy and list the constraints. In Section III
we discuss methods for solving Anderson impurity model based on the
slave--boson Gutzwiller and the Hubbard I approximations which can be used
to find these constraints. We also present our numerical comparisons with
the QMC data between various quantities such as the number of particles and
the quasiparticle residue. In Section IV we discuss the interpolational
approach and compare the self--energies and the Green functions with the QMC
results. Section V is the conclusion.

\section{Interpolative Approach}

To be specific, we concentrate on the Anderson impurity Hamiltonian 
\begin{eqnarray}
H &=&\sum_{\alpha =1}^{N}\epsilon _{f\alpha }f_{\alpha }^{+}f_{\alpha }+%
\frac{1}{2}\sum_{\alpha \beta }^{N}U_{\alpha \beta }n_{\alpha }^{f}n_{\beta
}^{f}+\sum_{\mathbf{k}\alpha }E_{\mathbf{k}\alpha }c_{\mathbf{k}\alpha
}^{+}c_{\mathbf{k}\alpha }  \notag \\
&+&\sum_{\mathbf{k}\alpha }[V_{\alpha }^{\ast }(\mathbf{k})f_{\alpha }^{+}c_{%
\mathbf{k}\alpha }+V_{\alpha }(\mathbf{k})c_{\mathbf{k}\alpha }^{+}f_{\alpha
}],  \label{HAM}
\end{eqnarray}%
describing the interaction of the impurity levels $\epsilon _{f\alpha }$
with bands of conduction electrons $E_{\mathbf{k}\alpha }$ via hybridization 
$V_{\alpha }(\mathbf{k})$. $U_{\alpha \beta }$ is the Coulomb repulsion
between different orbitals in the $f$--band. Inspired by the success of the
iterative perturbation theory \cite{DMFT}, in order to solve the Anderson
impurity model in general multiorbital case, we use a rational interpolative
formula for the self--energy. This can be either encoded into a simple form
of pole expansion%
\begin{equation}
\Sigma _{\alpha }(i\omega )=\Sigma _{\alpha }(i\infty )+\sum\limits_{n}{}%
\frac{W_{n}^{\alpha }}{i\omega -P_{n}^{\alpha }},  \label{RAT}
\end{equation}%
or, alternatively, in a form of continuous fraction expansion%
\begin{equation}
{\Sigma }_{\alpha }(i\omega )={\Sigma }_{\alpha }(i\infty )+\frac{A_{\alpha }%
}{i\omega -B_{\alpha }-\frac{C_{\alpha }}{i\omega -D_{\alpha }}...}
\label{CFR}
\end{equation}%
The coefficients $W_{n}^{\alpha }$, $P_{n}^{\alpha }$, in Eq. (\ref{RAT}) or
the coefficients $A_{\alpha },$ $B_{\alpha },C_{\alpha }$, ... in Eq. (\ref%
{CFR}) are to be determined together with the Hartree Fock value of the
self--energy ${\Sigma }_{\alpha }(i\infty )$. In principle, both
representations are equivalent and there is a well--defined non--linear
relationship between the parameters entering (\ref{RAT}) and (\ref{CFR}).

Our basic assumption is that only a few poles in the rational representation
(\ref{RAT}) or a few coefficients in continuous fraction expansion (\ref{CFR}%
) is necessary to reproduce an overall frequency dependence of the
self--energy necessary for the accuracies of the total energy calculations.
Extensive experience gained from solving Hubbard and periodic Anderson model
within DMFT at various ratios of the on--site Coulomb interaction $U$ to the
bandwidth $W$ shows the appearance of lower and upper Hubbard bands as well
as renormalized quasiparticle peak in the spectrum of one--electron
excitations \cite{DMFT}. To describe such three--peak structure we expect
that either two-- or at most three--pole formulae for the self--energy will
suffice. Thus, four (six) unknown parameters need to be determined when
using two (three) pole interpolating formulae.

In order to fix the coefficients we can explore several constrains for the
self--energy:

\textit{a)} \textit{Hartree--Fock value.} In the limit $i\omega \rightarrow
i\infty $ the self--energy takes its Hartree--Fock form 
\begin{equation}
{\Sigma }_{\alpha }(i\infty )=\sum_{\beta }U_{\alpha \beta }\langle n_{\beta
}\rangle .  \label{INF}
\end{equation}

\textit{b)} \textit{Moments.} The subleading high frequency behavior of the
self energy can be used to determine some additional coefficients. In the
case of the three pole approximation, we determine coefficients $A_{\alpha }$%
,$B_{\alpha }$ and $C_{\alpha }$ from the first three moments of the
self--energy $\Sigma _{\alpha }^{(1)},$ $\Sigma _{\alpha }^{(2)},\Sigma
_{\alpha }^{(3)}$ as follows

\begin{equation}
A_{\alpha }=\Sigma _{\alpha }^{(1)},  \label{AA1}
\end{equation}%
\begin{equation}
B_{\alpha }=\frac{\Sigma _{\alpha }^{(2)}}{\Sigma _{\alpha }^{(1)}},
\label{BA1}
\end{equation}%
\begin{equation}
C_{\alpha }=\frac{\Sigma _{\alpha }^{(3)}\Sigma _{\alpha }^{(1)}-(\Sigma
_{\alpha }^{(2)})^{2}}{(\Sigma _{\alpha }^{(1)})^{2}},  \label{CA1}
\end{equation}%
while in the two pole approximation only coefficient $A_{\alpha }$ should be
determined from the high frequency moment, as stated in Eq.~(\ref{AA1})
above. The self--energy moments can be expressed by the occupancies and
correlation functions that can be calculated within an approximation like
slave--boson Gutzwiller or Hubard I. The expressions for the self--energy
moments are worked out explicitely in the Appendix A.

\textit{c) Mass Renormalization. }If independent determination of the
quasiparticle residue, $Z_{\alpha }$, exists, the following relationship
holds 
\begin{equation}
\frac{{\partial \Sigma _{\alpha }}}{{\partial }i\omega }\mid _{i\omega
=0}=1-Z_{\alpha }^{-1}.  \label{MAS}
\end{equation}

\textit{d) Number of particles.} Sum over all Matsubara frequencies of the
Green's function gives number of particles, i.e.%
\begin{equation}
n_{\alpha }=T\sum_{i\omega }G_{f\alpha }(i\omega )\mathrm{e}^{i\omega 0^{+}},
\label{DEN}
\end{equation}%
where 
\begin{equation}
G_{f\alpha }(i\omega )=\frac{1}{i\omega -\epsilon _{f\alpha }-\Delta
_{\alpha }(i\omega )-\Sigma _{\alpha }(i\omega )}  \label{IMP}
\end{equation}%
defines the impurity Green function and $\Delta _{\alpha }(i\omega _{n})$ is
the hybridization function.

\textit{e)} \textit{Friedel sum rule.}

This is a relation between the total density and the real part of the
self--energy at zero frequency. It determines zero--frequency value of the
self--energy 
\begin{eqnarray}
n_{\alpha } &=&\frac{1}{2}+\frac{1}{\pi }\mathrm{arctg}\left( \frac{\epsilon
_{f\alpha }+\Re \Sigma _{\alpha }(i0^{+})+\Re \Delta _{\alpha }(i0^{+})}{\Im
\Delta _{\alpha }(i0^{+})}\right)  \notag \\
&+&\int\limits_{-i\infty }^{+i\infty }\frac{dz}{2\pi i}G_{f\alpha }(z)\frac{%
\partial \Delta _{\alpha }(z)}{\partial z}e^{z0^{+}}.  \label{FRD}
\end{eqnarray}

Formally, some of the constrains [(\textit{c}) and (\textit{e})] hold for
zero temperature only but we expect no significant deviations as long as we
stay at low temperatures.

We thus see that the interpolational scheme is defined completely once a
prescription for obtaining parameters such as $Z_{\alpha },$ $n_{\alpha }$
and $\Sigma _{\alpha }^{(n)}$ is given. For this purpose we will test two
popular methods: slave--boson Gutzwiller method \cite{Gutz} as described by
Kotliar and Ruckenstein \cite{Ruck} and the well--known Hubbard I
approximation \cite{Hubbard1}. We compare these results against more
accurate but computationally demanding Quantum Monte Carlo \cite{DMFT}
calculations.

Notice that once the parameters such as $Z_{\alpha }$ are computed from a
given approximate method, some of the quantities such as the total number of
particles, $n_{\alpha }$, and the value of the self--energy at zero
frequency, $\Sigma _{\alpha }(i0),$ can be computed fully
self--consistently. They can be compared with their non--self--consistent
values. If the approximate scheme already provides a good approximation for $%
n_{\alpha }$ and satisfies the Friedel sum rule, the self--consistency can
be avoided hence accelerating the calculation. Indeed we found that
inclusion of the self--consistency improves the Gutzwiller results in the
vicinity of the Mott transition only where the Hubbard bands get
increasingly important.

We now give the description of our approximate algorithms and then present
the comparison with the QMC calculations.

\section{Methods for Solving Impurity Model}

\subsection{\textit{Quantum Monte Carlo Method}}

The Quantum Monte Carlo method is a powerful and manifestly not perturbative
approach in either interaction $U$ or the bandwidth $W$. In the QMC method
one introduces a Hubbard--Stratonovich field and averages over it using the
Monte Carlo sampling. This is a controlled approximation using different
expansion parameter, the size of the mesh for the imaginary time
discretization. Unfortunately it is computationally very expensive as the
number of time slices and the number of Hubbard--Stratonovich fields
increases. Extensive description of this method can be found in Ref. %
\onlinecite{DMFT}. We will use this method to benchmark our calculations
with approximate but much faster algorithms described below.

\subsection{\textit{Slave--Boson Approach}}

A fast approach to solve a general impurity problem is the slave--boson
method\ \cite{Ruck, Fleszar,Hasegawa}. At the mean field level, it gives the
results similar to the famous Gutzwiller approximation \cite{Gutz}. However,
it is improvable by performing fluctuations around the saddle point. This
approach is accurate as it has been shown recently to give the exact
critical value of $U$ in the large degeneracy limit\ at half--filling \cite%
{Florens}.

The main idea is to rewrite atomic states consisting of $n$ electrons $%
\left\vert \gamma _{1},...,\gamma _{n}\right\rangle $, $0\leq n\leq N$ with
help of a set of slave--bosons $\{\psi _{n}^{\gamma _{1,...,}\gamma _{n}}\}$%
. In the following, we assume $SU(N)$ symmetric case, i.e., equivalence
between different states $\left\vert \gamma _{1},...,\gamma
_{n}\right\rangle $ for fixed $n$. Formulae corresponding to a more general
crystal--field case are given in Appendix B. The creation operator of a
physical electron is expressed via slave particles in the standard manner\ 
\cite{Fleszar}. In order to recover the correct non--interacting limit at
the mean--field level, the Bose fields $\psi _{n\text{ }}$ can be considered
as classical values found from minimizing a Lagrangian $L\{\psi _{n}\}$
corresponding to the Hamiltonian (\ref{HAM}). Two Lagrange multipliers $%
\lambda $ and $\Lambda $ should be introduced in this way, which correspond
to the following two constrains: 
\begin{eqnarray}
&&\sum_{n=0}^{N}C_{n}^{N}\psi _{n}^{2}=1,  \label{ONE} \\
&&\sum_{n=0}^{N}nC_{n}^{N}\psi _{n}^{2}=TN\sum_{i\omega }G_{g}(i\omega
)e^{i\omega 0^{+}}=\bar{n}.  \label{TWO}
\end{eqnarray}

The physical meaning of the first constrain is that the sum of probabilities
to find atom in any state is equal to one, and the second constrain gives
the mean number of electrons coinciding with that found from $G_{g}(i\omega
)= (i\omega -\lambda -b^{2}\Delta (i\omega ))^{-1}$. A combinatorial factor $%
C_{n}^{N}=\frac{N!}{n!(N-n)!}$ arrives due to assumed equivalence of all
states with $n$ electrons.

Minimization of $L\{\psi _{n}\}$ with respect to $\psi _{n}$ leads us to the
following set of equations to determine the quantities $\psi _{n}$: 
\begin{equation*}
\lbrack E_{n}+\Lambda -n\lambda ]\psi _{n}+nbT\sum_{i\omega }\Delta (i\omega
)G_{g}(i\omega )[LR\psi _{n-1}+\psi _{n}bL^{2}]+
\end{equation*}%
\begin{equation}
+(N-n)bT\sum_{i\omega }\Delta (i\omega )G_{g}(i\omega )[R^{2}b\psi
_{n}+LR\psi _{n+1}]=0,  \label{GUZ}
\end{equation}%
where $b=RL\sum_{n=1}^{N}C_{n-1}^{N-1}\psi _{n}\psi _{n-1}$, determines the
mass renormalization, and the coefficients $L=(1-\sum_{n=1}^{N}C_{n-1}^{N-1}%
\psi _{n}^{2})^{-1/2}$, $R=(1-\sum_{n=0}^{N}C_{n}^{N-1}\psi _{n}^{2})^{-1/2}$
are normalization constants as in Refs. \onlinecite{Ruck, Fleszar}. $%
E_{n}=\epsilon _{f}n+Un(n-1)/2$ is the total energy of the atom with $n$
electrons within $SU(N)$ approximation.

Eq.~(\ref{GUZ}), along with the constrains (\ref{ONE}), (\ref{TWO})
constitute a set of non--linear equations which have to be solved
iteratively. In practice, we consider Eq. (\ref{GUZ}) as an eigenvalue
problem with $\Lambda $ being the eigenvalue and $\psi _{n}$ being the
eigenvectors of the matrix. The physical root corresponds to the lowest
eigenvalue of $\Lambda $ which gives a set of $\psi _{n}$ determining the
mass renormalization $Z=b^{2}.$ Since the matrix to be diagonalized depends
non--linearly on $\psi _{n}$ via the parameters $L,R,$ and $b$ and also on $%
\lambda $, the solution of the whole problem assumes the self--consistency:
(i) we build an initial approximation to $\psi _{n}$ (for example the
Hartree--Fock solution) and fix some $\lambda $, (ii) we solve eigenvalue
problem and find new normalized $\psi _{n}$, (iii) we mix new $\psi _{n}$
with the old ones using the Broyden method~\cite{Broyden} and build new $%
L,\,R,$ and $b$. Steps (ii) and (iii) are repeated until the
self--consistency with respect to $\psi _{n}$ is reached. During the
iterations we also vary $\lambda $ to obey the constrains. The described
procedure provides a stable computational algorithm for solving AIM and
gives us an access to the low--frequency Green's function and the
self--energy of the problem via knowledge of the slope of $\Im \Sigma
(i\omega )$ and the value $\Re \Sigma (0)$ at zero frequency.

The described slave--boson method gives the following expression for the
self--energy: 
\begin{equation}
\Sigma (i\omega )=(1-b^{-2})i\omega -\epsilon _{f}+\lambda b^{-2}.
\label{SIG}
\end{equation}%
The impurity Green function $G_{f}(i\omega )$ in this limit is given by the
expression 
\begin{equation}
G_{f}(i\omega )=b^{2}G_{g}(i\omega ).  \label{GGF}
\end{equation}

As an illustration, we now give the solution of Eq. (\ref{GUZ}) for
non--degenerate case ($N=2)$ and at the particle--hole symmetry point with $%
\epsilon _{imp}-\mu =-\frac{U}{2}(N-1)$. Consider a dynamical mean--field
theory for the Hubbard model which reduces the problem to solving the
impurity model subject to the self--consistency condition with respect to $%
\Delta (i\omega )$. Starting with the semicircular density of states (DOS),
the self--consistency condition is given by $\Delta (i\omega
)=t^{2}G_{f}\left( i\omega \right) $ with parameter $t$ being the quarter of
bandwidth. We obtain the following simplifications: $L=R=\sqrt{2}$, $\lambda
=0$, $\psi _{0}=\psi _{2},b=4\psi _{1}\psi _{2}$ and $G_{g}(i\omega
)=[i\omega -t^{2}b^{2}G_{g}(i\omega )]^{-1}.$ The sum $T\sum_{i\omega
}\Delta (i\omega )G_{g}(i\omega )$ appeared in Eq.~(\ref{GUZ}) scales as $%
2t\alpha $ with the constant $\alpha $ being the characteristic of a
particular density of states and approximately equal to --0.2 in the
semicircular DOS case. The self--consistent solution of Eq. (\ref{GUZ}) is
therefore possible and simply gives $\psi _{2}^{2}=\frac{U}{128t\alpha }+%
\frac{1}{4}$. The Mott transition occurs when no sites with double
occupancies can be found, i.e. when $\psi _{0}=\psi _{2}=0.$ A critical
value of $U_{c}=32t|\alpha |=8W|\alpha |$ with $W$ being the bandwidth. For $%
\alpha \approx -0.2$, this gives $U_{c}\approx 1.6W$ and reproduces the
exact result for $U_{c}=1.49W$ within a few percent accuracy. As degeneracy
increases, critical $U$ is shifted towards higher values. From numerical
calculations we obtained the following values of the critical interactions
in the half--filled case $U_{c}\approx 3W$ for $N=6$ ($p$--level)$,$ $%
U_{c}\approx 4.5W$ for $N=10$ ($d$--level)$,$ and $U_{c}\approx 6W$ for $%
N=14 $ ($f$--level).

Density--density correlation function $\langle n_{\alpha }n_{\beta }\rangle $
(where $\alpha \neq \beta $) for local states with $n$ electrons is
proportional to the number of pairs formed by $n$ particles $%
C_{2}^{n}/C_{2}^{N}$. Since the probability for $n$ electrons to be occupied
is given by: $P_{n}=\psi _{n}^{2}C_{n}^{N}$, the physical density--density
correlator can be deduced from: $\langle nn\rangle
=\sum_{n}C_{2}^{n}/C_{2}^{N}P_{n}$. Similarly, the triple occupancy can be
calculated from: $\langle nnn\rangle =\sum_{n}C_{3}^{n}/C_{3}^{N}P_{n}$.

Let us now check the accuracy of this method by comparing it's results with
the QMC data. We consider the two--band Hubbard model within $SU(4)$
orbitally degenerate case. Hybridization $\Delta (i\omega )=\sum_{\mathbf{k}%
}V(\mathbf{k})/(i\omega -E_{\mathbf{k}})$ satisfies the DMFT\
self--consistency condition of the Hubbard model on a Bethe lattice%
\begin{equation}
\Delta (i\omega )=t^{2}G(i\omega ),  \label{SCF}
\end{equation}%
where $t$ is a quarter of the bandwidth $W$. The Coulomb interaction is
chosen to be $U=4d$ ($d$ is the half--bandwidth) which is sufficiently large
to open the Mott gap at integer fillings. All calculations are done for the
temperature $T=1/16d$.

\begin{figure}[h]
\includegraphics[height=0.5\textwidth]{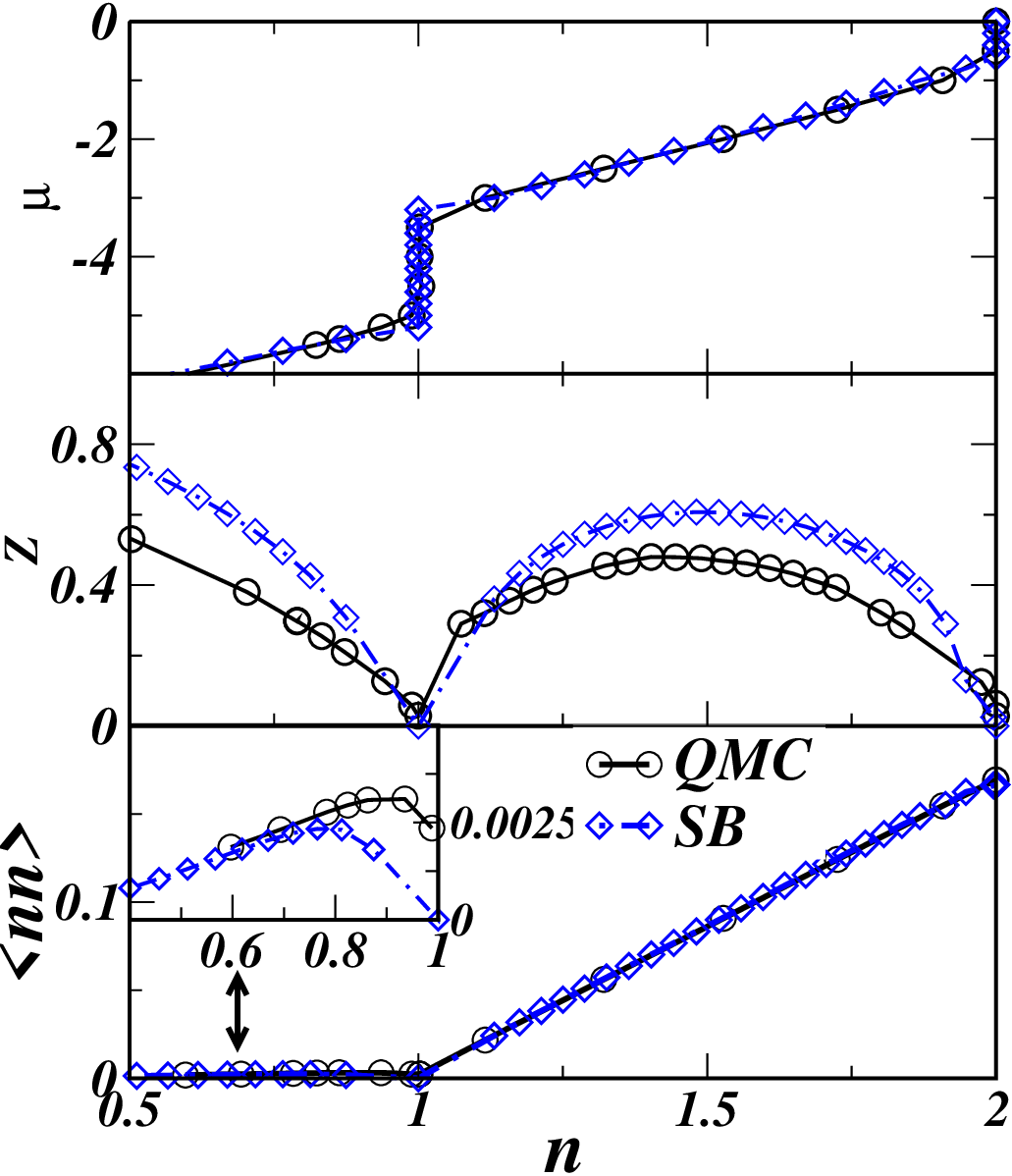}\\[-0.3cm]
\caption{Comparison between the slave boson mean field and the QMC
calculation for (a) concentration versus chemical potential, (b) dependence
of the spectral weight $Z$ on concentration, and (c) density--density
correlation function, $\langle n_{\protect\alpha }n_{\protect\alpha ^{\prime
}}\rangle $ versus filling, $\bar{n}$, in the two--band Hubbard model.}
\label{Fig1}
\end{figure}

We first compare the average number of electrons vs. chemical potential
determined from the slave bosons which is plotted in Fig.~\ref{Fig1}(a).
This quantity is sensitive to the low--frequency part of the Green function
which should be described well by the present method. We see that it
reproduces the QMC data with a very high accuracy. 
The quasiparticle residue $Z=1/(1-\frac{\partial \Im \Sigma (i\omega )}{%
\partial i\omega })|_{i\omega \rightarrow 0}$ versus filling $\bar{n}$ is
plotted in Fig. \ref{Fig1}(b). The slave--boson method gives the Fermi
liquid and provides a good estimate for the quasiparticle residue. Indeed,
the agreement with the QMC results is satisfactory. Also, recent findings 
\cite{Florens} suggest that at half--filling $Z$ deduced from slave bosons
becomes exact when $N\rightarrow \infty $. Still the discrepancy of the
order of 30\% can be found. In fact, we have performed several additional
calculations for other degeneracies ($N=2$ and $6$) and for various
parameters regimes. The trend to underestimate mass renormalization by about
30\% can be seen in almost all cases (it disappears only when $U$ approaches
zero). This empirical finding is useful as we can use the reduced values of $%
Z$ needed for our interpolating formula.

Fig.~\ref{Fig1}(c) shows the density--density correlation function $\langle
n_{\alpha }n_{\alpha ^{\prime }}\rangle $ as a function of average
occupation $\bar{n}$. The discrepancy is most pronounced for fillings $\bar{n%
}<1$ [see the inset of Fig. \ref{Fig1}(c)] where the absolute values of $%
\langle n_{\alpha }n_{\alpha ^{\prime }}\rangle $ are rather small. Although
our slave--boson technique captures only the quasiparticle peak, it gives
the correlation function in reasonable agreement with the QMC for dopings
not too close to the Mott transition.

\subsection{\textit{Hubbard I Approximation}}

Now we turn to the Hubbard\ I approximation\ \cite{Hubbard1} which is
closely related to the moments expansion method\ \cite{Moments}. One
introduces Hubbard operators and the corresponding Green function $%
G_{nm}(i\omega )$ defined for these Hubbard operators. Within $SU(N)$ we are
summing over certain set of Hubbard operators which induce transitions
between $n$--times and $n+1$--times occupied local level. The impurity Green
function is 
\begin{equation}
G_{f}(i\omega )=\sum_{nm}G_{nm}(i\omega ),  \label{HUB}
\end{equation}%
where the matrix $[G_{nm}(i\omega )]^{-1}=[G_{nm}^{at}(i\omega
)]^{-1}-\Delta (i\omega ),n,m=0,N-1$. $G_{nm}^{at}(i\omega )$ is the atomic
Green function 
\begin{equation}
G_{nm}^{at}(i\omega )=\delta _{nm}\frac{C_{n}^{N-1}(X_{n}+X_{n+1})}{i\omega
+\mu -E_{n+1}+E_{n}}.  \label{GAT}
\end{equation}%
The coefficients $X_{n}$ are the probabilities to find atom with $n$
electrons similar to the coefficients $\psi _{n}^{2}$ introduced above. They
can be also used to find the averages $\langle nn\rangle $ in the same way.
These numbers are normalized to unity, $\sum_{n=0}^{N}C_{n}^{N}X_{n}=%
\sum_{n=0}^{N-1}C_{n}^{N-1}(X_{n}+X_{n+1})=1,$ and are expressed via
diagonal elements of $G_{nm}(i\omega )$ as follows: $X_{n}=-T\sum_{i\omega
}G_{nn}(i\omega )e^{-i\omega 0^{+}}/C_{n}^{N-1}$. The mean number of
electrons can be measured as follows: $\bar{n}=\sum_{n=0}^{N}nC_{n}^{N}X_{n}$
or as: $\bar{n}=TN\sum_{i\omega }G_{f}(i\omega )e^{i\omega 0^{+}}.$

Note that when $\Delta (i\omega )\equiv 0$, $G_{f}(i\omega )$ is reduced to $%
\sum_{nm}G_{nm}^{at}(i\omega ),$ i.e. Hubbard\ I reproduces the atomic
limit. Setting $U\equiv 0$ gives $G_{f}(i\omega )=[i\omega +\mu -\epsilon
_{f}-\Delta (i\omega )]^{-1}$, which is the correct band limit.
Unfortunately, at half--filling this limit has a pathology connected to the
instability towards Mott transition at any interaction strength $U$. To see
this, we consider a dynamical mean--field theory for the Hubbard model.
Using semicircular density of states, we obtain $G_{f}(i\omega
)=[1-t^{2}G_{f}\left( i\omega \right) G_{at}(i\omega )]^{-1}G_{at}(i\omega )$
and conclude that for any small $U$ $\ $the system opens a pathological gap
in the spectrum. Clearly, using Hubbard I only, the behavior of the Green
function at $i\omega \rightarrow 0$ cannot be reproduced. This emphasizes
the importance of using the slave--boson treatment at small frequencies.

\begin{figure}[h]
\includegraphics[height=0.5\textwidth]{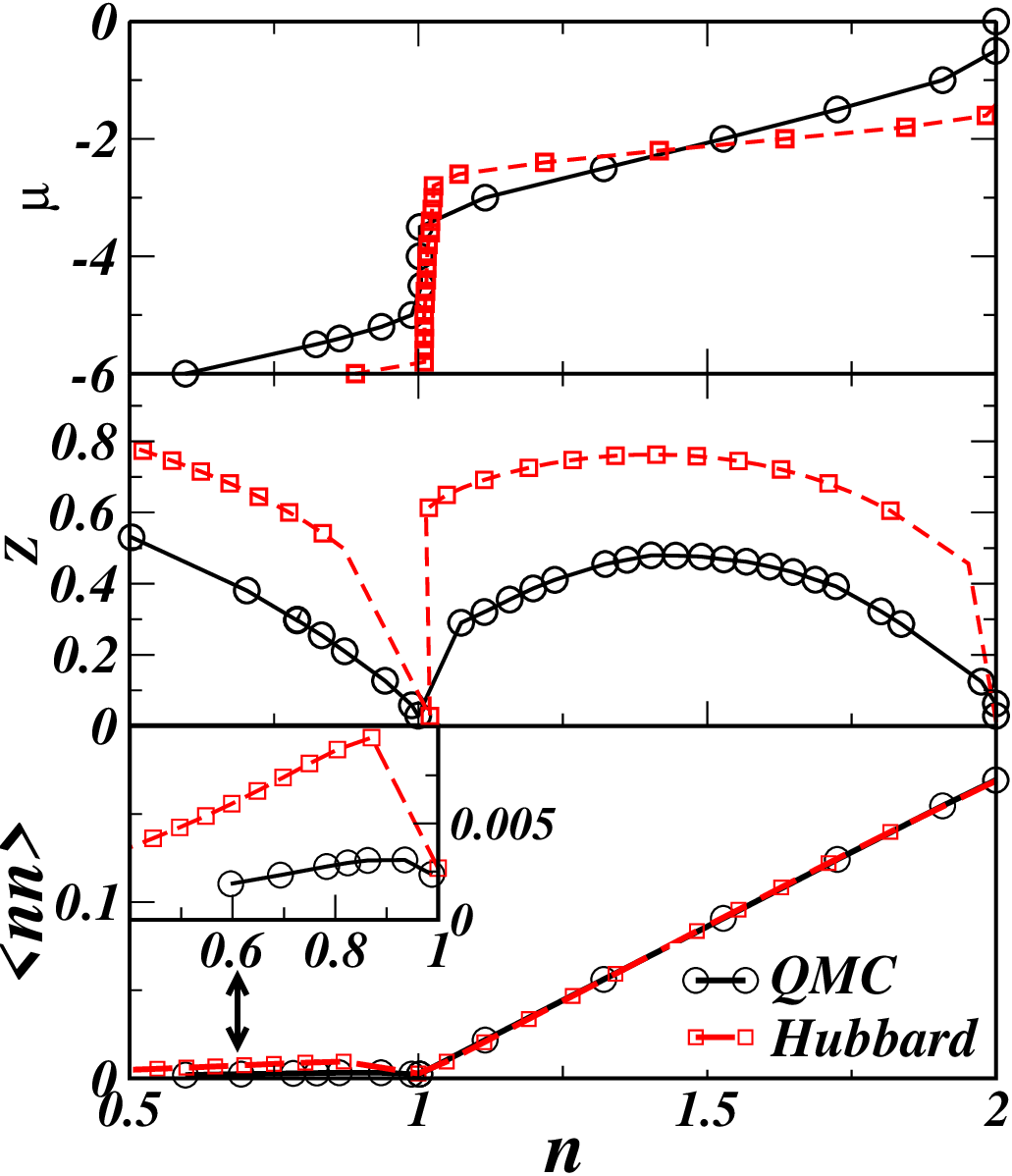}\\[-0.3cm]
\caption{Comparison between the Hubbard I and the QMC calculation for (a)
concentration versus chemical potential, (b) dependence of the spectral
weight $Z$ on concentration, and (c) density--density correlation function, $%
\langle n_{\protect\alpha }n_{\protect\alpha ^{\prime }}\rangle $ versus
filling, $\bar{n}$, in the two--band Hubbard model.}
\label{Fig2}
\end{figure}

Let us now check the accuracy of this method against QMC. We again consider
the two--band Hubbard model within $SU(4)$ symmetry as in the case of the
slave--boson method described earlier. The $\bar{n}(\mu )$ is plotted in
Fig.~\ref{Fig2}(a). The Hubbard I approximation does not give satisfactory
agreement with the QMC data because it misses the correct behavior at low
frequencies. 
The comparisons for $Z(\bar{n})$ is plotted in Fig. \ref{Fig2}(b). These
numbers are extracted from the low--frequency slope of the imaginary part of
the self--energy on the imaginary axis. Surprizingly, the Hubbard I $%
Z^{\prime }$s show a relatively good behavior when comparing to the QMC
method. However, the patology of this approximation at half--filling would
predict $Z=0$ for any $U$, which is a serious warning not to use it for
extracting the quasiparticle weight. Fig.~\ref{Fig2}(c) shows $\langle
nn\rangle $ as a function of average occupation $\bar{n}$. As this quantity
is directly related to the high--frequency expansion one may expect a better
accuracy here. However, comparing Fig. \ref{Fig2}(c) and Fig. \ref{Fig1}(c),
it is clear that the slave boson method gives more accurate double
occupancy. This is due to the fact that the density matrix obtained by the
slave boson method is of higher quality than the one obtained from the
Hubbard I approximation.

The results of these calculations suggest that the slave--boson Gutzwiller
method shows a good accuracy in comparison with the QMC. Therefore it can be
used to determine the unknown coefficients in the interpolational form of
the self--energy, Eqs. (\ref{RAT}) or (\ref{CFR}). At the same time the use
of the Hubbard approximation should be avoided. Interestingly, while more
sophisticated QMC approach captures both the quasiparticle peak and the
Hubbard bands this is not the case for the slave--boson mean field method.
To obtain the Hubbard bands in this method fluctuations need to be computed,
which would be very tedious in a general multiorbital situation. However the
slave--boson method delivers many parameters in a good agreement with the
QMC results, and hence it can be used to obtain the parameters that enter
the rational approximation. In this way slave--boson approach generates a
better Green function with quasiparticles and Hubbard bands \textit{at no
extra computational cost}.

\section{\textbf{Results of the Interpolative Scheme}}

We now turn to the comparison of the self--energie obtained using the
formula (\ref{CFR}) and the corresponding Green functions after the formula (%
\ref{IMP}) against the predictions of the Quantum Monte Carlo method. We
have studied the results of the two--pole and the three--pole interpolations
which require the determination of either 4 or 6 coefficients. The
interpolational formula with two--poles needs finding the values of $\Sigma
(0)$, $Z$ and the first high--frequency moment of the self--energy $\Sigma
^{(1)}.$ The condition that the total number of electrons found from the
Green function (\ref{IMP}) is the same as predicted by the slave--boson
method can be used to fix the fourth coefficient. To explore the three--pole
interpolation, two more conditions are required. We have used additional
high frequency moments $\Sigma ^{(2)},$ and $\Sigma ^{(3)}$ which can be
determined if one computes correlational functions $\langle nn\rangle $, $%
\langle nnn\rangle $, $\langle nnnn\rangle $. The precise functional form of
these moments is given in Appendix A. 
\begin{figure}[h]
\includegraphics[height=0.32\textwidth]{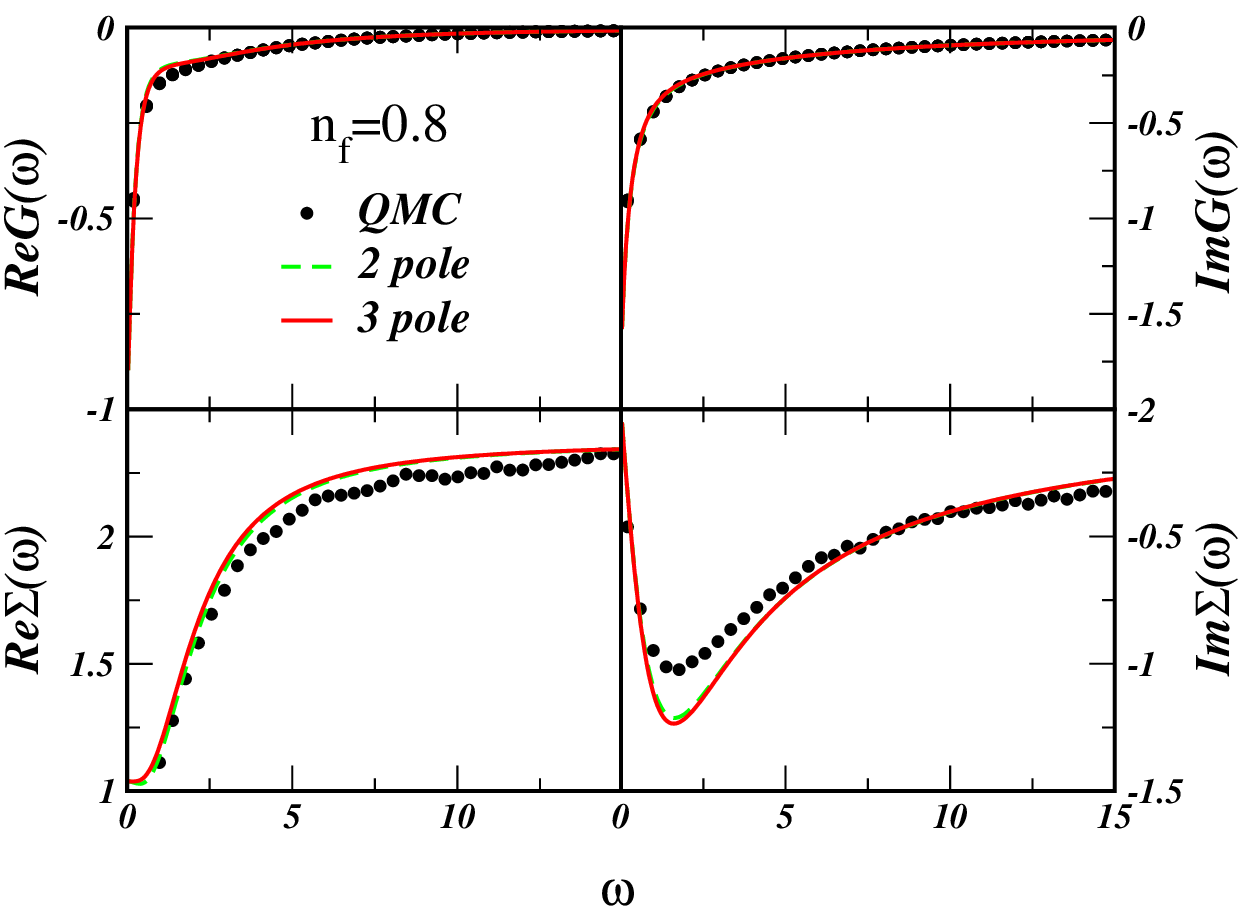}
\caption{Comparison between real and imaginary parts of the Green function
and the self--energy obtained from the interpoaltive method and the Quantum
Monte Carlo calculation for the two--band Hubbard model at filling $\bar{n}%
=0.8$.}
\label{Fig3}
\end{figure}

We perform the comparison of the interpolational scheme against QMC\ for two
characteristic values of the concentrations: $\bar{n}=0.8$ representing the
case $\bar{n}<1$ and $\bar{n}=1.2$ representing the case $\bar{n}>1$. Both
cases are chosen to be close to the Mott transition. This is a hard test for
our approach as we have found that the predictions of the interpolative
scheme and the QMC data are practically indistinguishable in the metallic
region far from the Mott transition ($U\lesssim W$). The two--band Hubbard
model within $SU(4)$ symmetry, $U=4d$, and $\Delta (i\omega)=t^{2}G(i\omega
) $ self--consistency condition is used. 
\begin{figure}[h]
\includegraphics[height=0.32\textwidth]{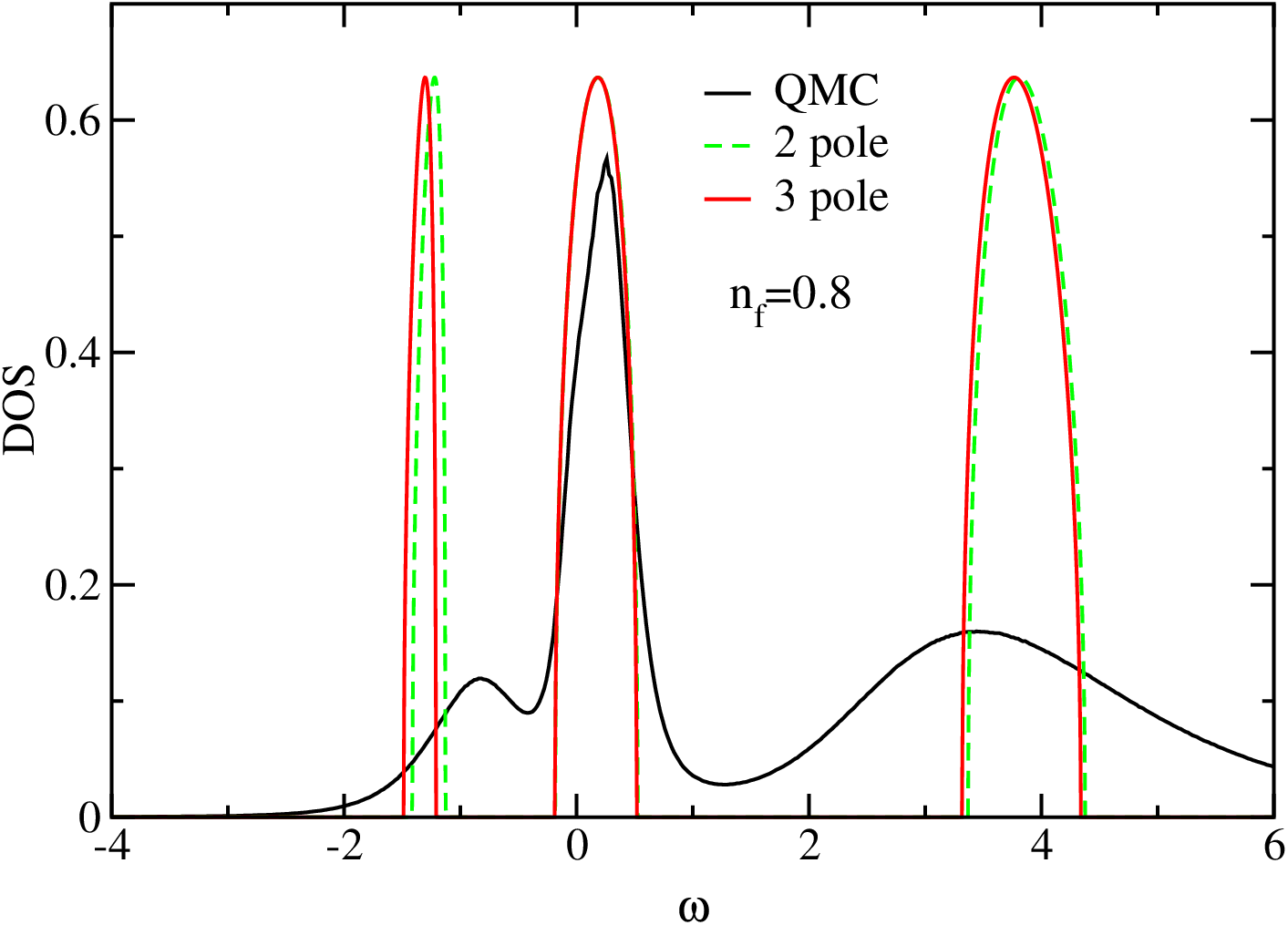}\\[-0.3cm]
\caption{Comparison between\ the one--electron densities of states obtained
from interpolative formula and the Quantum Monte Carlo calculationfor the
two--band Hubbard model at filling $\bar{n}=0.8$.}
\label{Fig4}
\end{figure}

Fig.~\ref{Fig3} shows the comparison of the real and imaginary parts of the
self--energy and the Green function obtained by the two-- and three--pole
interpolations with the QMC results. Both interpolational schemes give
pretty good agreement at all frequencies. These results are obtained with
the values of mass renormalizations $Z$ reduced by 30\% which gives a better
agreement with $Z_{QMC}$ as we discussed in Section III. The results with
bare $Z^{\prime }s$ deduced from the slave--boson method are also found to
be quite accurate but not as good as we advertise in Fig.~\ref{Fig3}. We
have also performed the comparison for the one--electron density of states
which are shown in Fig.~\ref{Fig4}. One can see the appearance of the lower
and upper Hubbard bands as well as renormalized quasiparticle peak. Note a
complete match in the positions of the Hubbard band as well as correct
bandwidth for the quasiparticles as compared to the Quantum Monte Carlo
calculations. To obtain such remarkable agreement, the use of renormalized
values of $Z$ is essential. 
\begin{figure}[h]
\includegraphics[height=0.32\textwidth]{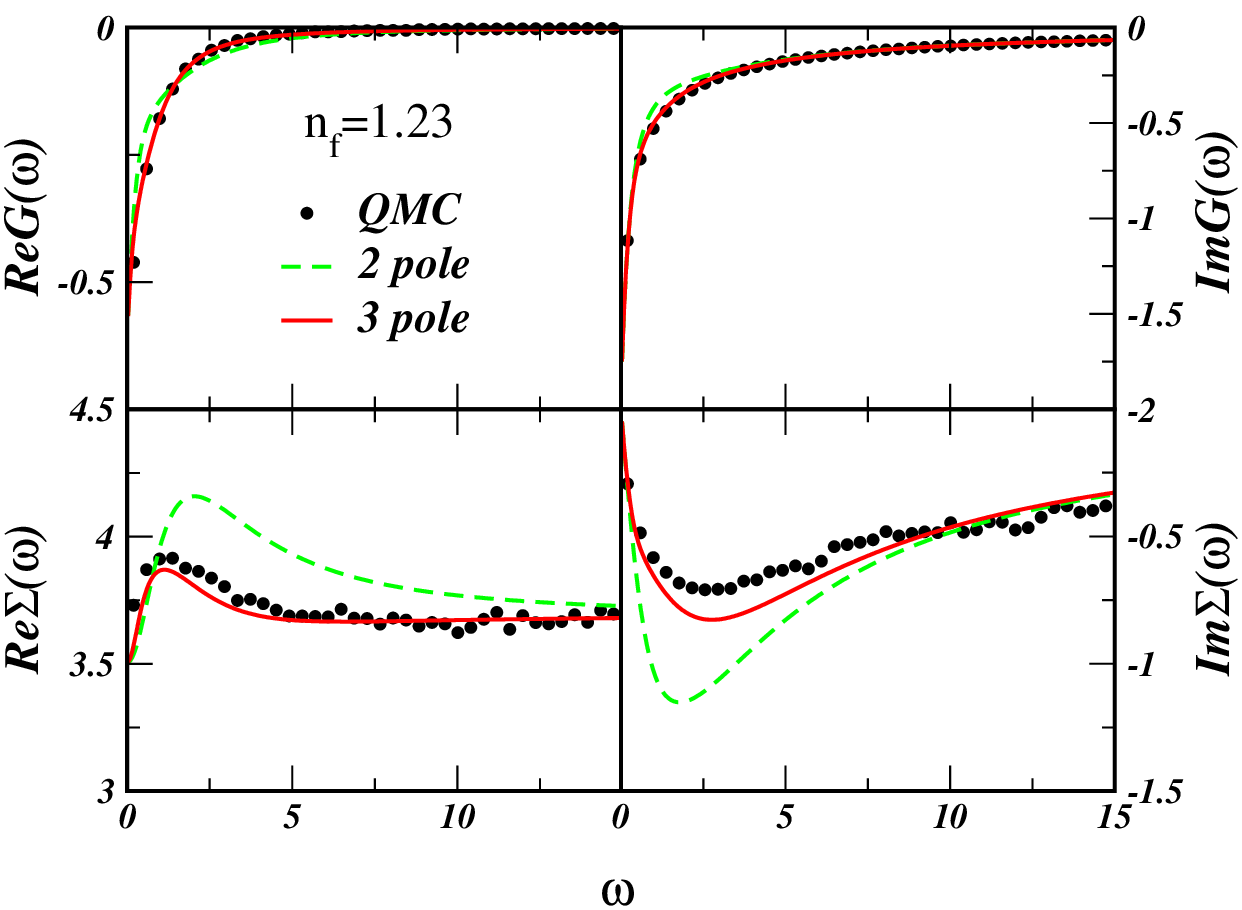}\\[-0.3cm]
\caption{Comparison between real and imaginary parts of the Green function
and the self--energy obtained from the interpoaltive method and the Quantum
Monte Carlo calculation for the two--band Hubbard model at filling $\bar{n}%
=1.23$.}
\label{Fig5}
\end{figure}

\begin{figure}[h]
\includegraphics[height=0.32\textwidth]{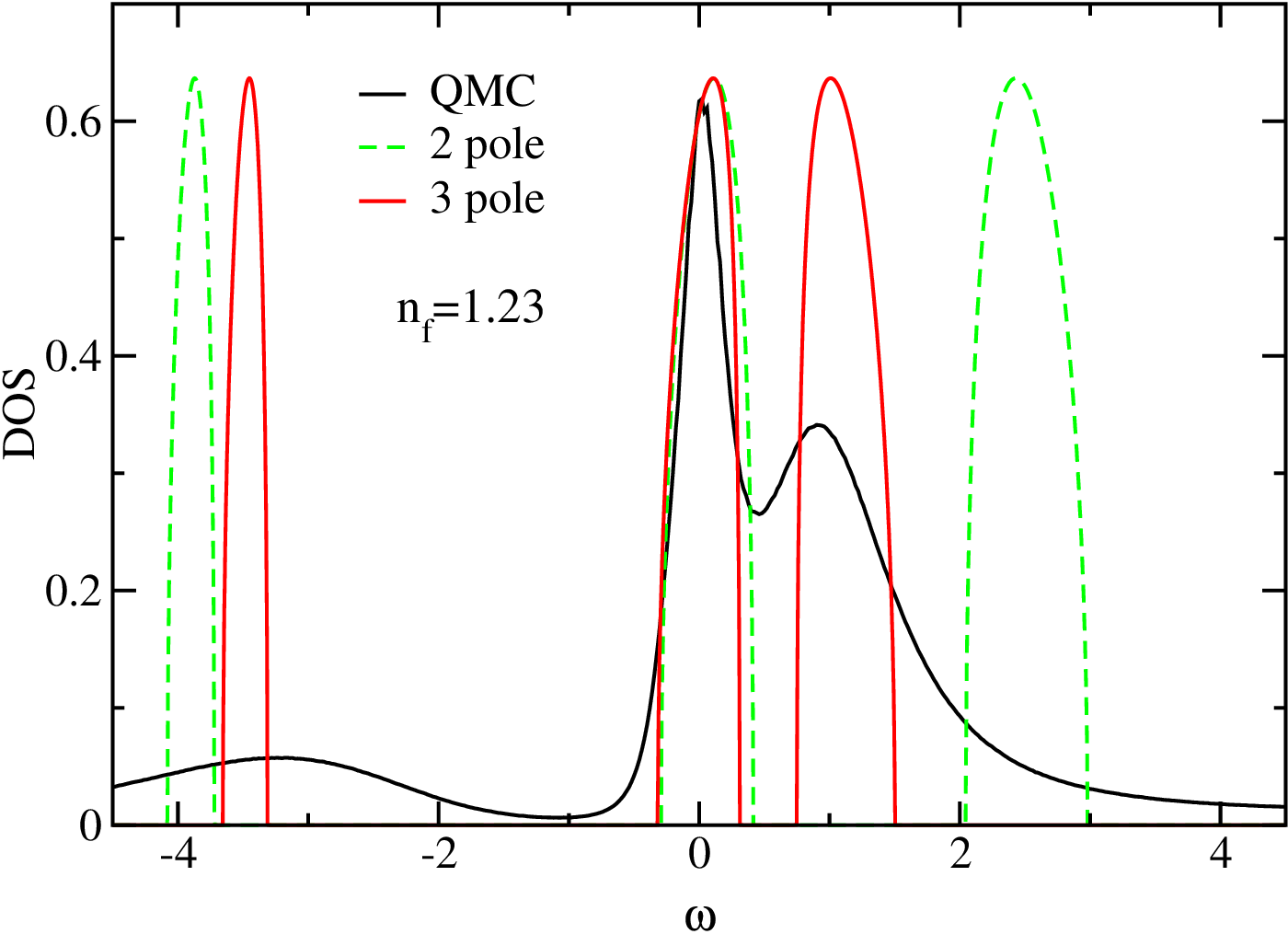}\\[-0.3cm]
\caption{Comparison between\ the one--electron densities of states obtained
from interpolative formula and the Quantum Monte Carlo calculationfor the
two--band Hubbard model at filling $\bar{n}=1.23$.}
\label{Fig6}
\end{figure}

In Fig.~\ref{Fig5}, we show our results for the case $\bar{n}=1.2$ using the
two-- and three--pole interpolational formula (\ref{CFR}). As far as the
two--pole scheme is concerned, the overall agreement is satisfactory but a
discrepancy exists in the region of intermediate imaginary frequencies. It
is directly related to the misplaced upper Hubard band as can be seen in
Fig.~\ref{Fig6}. This discrepancy disappears in the three--pole
interpolation as is evident from the Fig.~\ref{Fig5}. Comparison of the
one--electron density of states is shown in Fig.~\ref{Fig6}. Clearly, the
two--pole interpolation is inadequate in this regime, while the three--pole
interpolating scheme gives correct position of Hubbard bands and
quasiparticle peak, all basic features of the Anderson impurity problem.

Finally, we would like to comment on the renormalization of the
quasiparticle residue $Z$ beyond the slave boson mean--field result. While
we found empirically that the 30\% renormalization significantly improves
our fits to the QMC data, over a broad region of parameters, this cannot
hold everywhere since $Z$ should go to $1$ when $U$ approaches zero. The
reduction of $Z$ beyond the mean--field results, in the regions where the
correlations are strong, can be thought to arise from fluctuation
corrections of the slave boson around the saddle point value. A
computationally efficient way to evaluate these corrections, which will also
give rise to the quasiparticle lifetime not included in this paper, are
important open problems left for future work. 

\section{Conclusion}

To summarize, this paper shows the possibility to extract parameters for the
interpolating self--energy of the Anderson impurity model such as the
self--energy at zero frequency, the mass renormalization, the density and
the high frequency moments using computationally very efficient slave--boson
mean field approach. These data are then used to reconstruct the
self--energy at all frequencies using a rational interpolation formula. We
find that three--pole interpolation predicts the self--energy quite
accurately as compared to the results of more sophisticated approaches such
the Quantum Monte Carlo method.

As a general conclusion, we have thus found that interpolative formulae with
several poles are sufficient to fit the self--energies in a broad range of
the parameters of the quantum impurity problem and the slave--boson
mean--field method is capable to deliver all necessary ingredients for this
algorithm. The approach can be used for the evaluation of LDA+DMFT total
energies and is a very direct way to obtain spectra with a three--peak
structure, characteristic of the strongly correlated regime, with very
little computational effort.

The work was supported by NSF--DMR Grants 0096462, 02382188, 0312478,
0342290 and US DOE Grant No DE-FG02--99ER45761. The authors also acknowledge
the financial support from the Computational Material Science Network
operated by US DOE and from the Ministry of Education, Science and Sport of
Slovenia.

\begin{widetext}

\section{Appendix A}

High--frequency moment expansions for the Green function and the
self--energy can be worked out:
\begin{equation}
G_{\alpha }(i\omega )=\frac{G_{\alpha }^{(1)}}{\omega }+\frac{G_{\alpha
}^{(2)}}{\omega ^{2}}+\frac{G_{\alpha }^{(3)}}{\omega ^{3}}...,  \label{EXP}
\end{equation}%
where the moments of the Green function are given by the following formulae%
\begin{eqnarray}
G_{\alpha }^{(1)} &=&1,  \label{GA1} \\
G_{\alpha }^{(2)} &=&\epsilon _{f\alpha }+\sum_{\beta }U_{\alpha \beta
}\langle n_{\beta }\rangle ,  \label{GA2} \\
G_{\alpha }^{(3)} &=&\epsilon _{f\alpha }^{2}+2\epsilon _{f\alpha
}\sum_{\beta }U_{\alpha \beta }\langle n_{\beta }\rangle +\sum_{\beta \beta
^{\prime }}U_{\alpha \beta }U_{\alpha \beta ^{\prime }}\langle n_{\beta
}n_{\beta ^{\prime }}\rangle +\sum_{\mathbf{k}}|V_{\alpha }(\mathbf{k})|^{2},
\label{GA3} \\
G_{\alpha }^{(4)} &=&\epsilon _{f\alpha }^{3}+3\epsilon _{f\alpha
}^{2}\sum_{\beta }U_{\alpha \beta }\langle n_{\beta }\rangle +3\epsilon
_{f\alpha }\sum_{\beta \beta ^{\prime }}U_{\alpha \beta }U_{\alpha \beta
^{\prime }}\langle n_{\beta }n_{\beta ^{\prime }}\rangle +\sum_{\beta \beta
^{\prime }\beta ^{\prime \prime }}U_{\alpha \beta }U_{\alpha \beta ^{\prime
}}U_{\alpha \beta ^{\prime \prime }}\langle n_{\beta }n_{\beta ^{\prime
}}n_{\beta ^{\prime \prime }}\rangle ,  \notag \\
&+&\sum_{\mathbf{k}}|V_{\alpha }(\mathbf{k})|^{2}(E_{\mathbf{k}\alpha
}+2\epsilon _{f\alpha }+2\sum_{\beta }U_{\alpha \beta }\langle n_{\beta
}\rangle )+\sum_{\mathbf{k},\beta }U_{\alpha \beta }^{2}V_{\beta }(\mathbf{k}%
)\langle \left( 2n_{\alpha }-1\right) f_{\beta }^{\dagger }c_{\mathbf{k}%
\beta }\rangle .  \label{GA4}
\end{eqnarray}%
The expansion for the self--energy is obtained as follows%
\begin{equation}
\Sigma _{\alpha }(i\omega )=\omega +\mu -\Delta _{\alpha }(i\omega )-\left( 
\frac{G_{\alpha }^{(1)}}{\omega }+\frac{G_{\alpha }^{(2)}}{\omega ^{2}}+%
\frac{G_{\alpha }^{(3)}}{\omega ^{3}}+\cdots \right) ^{-1}=\Sigma _{\alpha
}(i\infty )+\frac{\Sigma _{\alpha }^{(1)}}{\omega }+\frac{\Sigma _{\alpha
}^{(2)}}{\omega ^{2}}+\frac{\Sigma _{\alpha }^{(3)}}{\omega ^{3}}...,
\label{SAM}
\end{equation}%
where the self--energy moments are given by
\begin{eqnarray}
\Sigma _{\alpha }^{(1)} &=&G_{\alpha }^{(3)}-[G_{\alpha
}^{(2)}]^{2}-\sum_{k}|V_{k}|^{2}=\sum_{\beta \beta ^{\prime }}U_{\alpha
\beta }U_{\alpha \beta ^{\prime }}(\langle n_{\beta }n_{\beta ^{\prime
}}\rangle -\langle n_{\beta }\rangle \langle n_{\beta ^{\prime }}\rangle ),
\label{SA1} \\
\Sigma _{\alpha }^{(2)} &=&G_{\alpha }^{(4)}-2G_{\alpha }^{(2)}G_{\alpha
}^{(3)}+[G_{\alpha }^{(2)}]^{3}-\sum_{\mathbf{k}}|V_{\alpha }(\mathbf{k}%
)|^{2}E_{\mathbf{k}\alpha }=  \notag \\
&=&\sum_{\beta \beta ^{\prime }\beta ^{\prime \prime }}U_{\alpha \beta
}U_{\alpha \beta ^{\prime }}U_{\alpha \beta ^{\prime \prime }}(\langle
n_{\beta }n_{\beta ^{\prime }}n_{\beta ^{\prime \prime }}\rangle -2\langle
n_{\beta }n_{\beta ^{\prime }}\rangle \langle n_{\beta ^{\prime \prime
}}\rangle +\langle n_{\beta }\rangle \langle n_{\beta ^{\prime }}\rangle
\langle n_{\beta ^{\prime \prime }}\rangle )  \notag \\
&+&\epsilon _{f\alpha }\sum_{\beta \beta ^{\prime }}U_{\alpha \beta
}U_{\alpha \beta ^{\prime }}(\langle n_{\beta }n_{\beta ^{\prime }}\rangle
-\langle n_{\beta }\rangle \langle n_{\beta ^{\prime }}\rangle )+\sum_{%
\mathbf{k},\beta }U_{\alpha \beta }^{2}V_{\beta }(\mathbf{k})\langle \left(
2n_{\alpha }-1\right) d_{\beta }^{\dagger }c_{\mathbf{k}\beta }\rangle ,
\label{SA2}\\
\Sigma_\a^{(3)} &=& G_\a^{(5)}-2 G_\a^{(4)}
G_\a^{(2)}-\left[G_\a^{(3)}\right]^2+3 G_\a^{(3)}
\left[G_\a^{(2)}\right]^2-\left[G_\a^{(2)}\right]^4-\sum_\vk|V_\a(\vk)|^2
E_{\vk\a}^2= \nonumber\\
&=&\sum_{\b_1\b_2\b_3\b_4}U_{\a\b_1}U_{\a\b_2}U_{\a\b_3}U_{\a\b_4}
\left(\langle
n_{\b_1}n_{\b_2}n_{\b_3}n_{\b_4}\rangle-2\langle
n_{\b_1}n_{\b_2}n_{\b_3}\rangle\langle n_{\b_4}\rangle-\langle
n_{\b_1}n_{\b_2}\rangle\langle n_{\b_3}n_{\b_4}\rangle
\right.
\nonumber\\&&\qquad\qquad\qquad\qquad\qquad\qquad
+ \left.
3\langle
n_{\b_1}n_{\b_2}\rangle\langle n_{\b_3}\rangle\langle n_{\b_4}\rangle-
\langle n_{\b_1}\rangle\langle n_{\b_2}\rangle\langle n_{\b_3}\rangle\langle n_{\b_4}\rangle
\right)\nonumber\\
&+& 2\epsilon_\a\sum_{\b_1\b_2\b_3}U_{\a\b_1}U_{\a\b_2}U_{\a\b_3}
\left(\langle n_{\b_1}n_{\b_2}n_{\b_3}\rangle-2\langle
n_{\b_1}n_{\b_2}\rangle\langle n_{\b_3}\rangle+\langle
n_{\b_1}\rangle\langle n_{\b_2}\rangle\langle n_{\b_3}\rangle
\right)\nonumber\\
&+&\epsilon_\a^2\sum_{\b_1\b_2}U_{\a\b_1}U_{\a\b_2}
\left(\langle n_{\b_1}n_{\b_2}\rangle-\langle
n_{\b_1}\rangle\langle n_{\b_2}\rangle\right)+{\cal N}_\a^{(3)}.
\label{SA3}
\end{eqnarray}
The nonlocal part of the self--energy moment, which explicitly
depends on hybridization matrix $V$ and vanishes in the atomic limit,
is explicitly quoted only for the second moment (last term in
Eq.~(\ref{SA2})). In the third order, the nonlocal term is denoted by
${\cal N}_\a^{(3)}$ and has not yet been evaluated. In practical
calculations, we neglected it's contribution.

Within $SU(N)$ symetric case the formulae are simplified. First, introduce
the definitions: 
\begin{eqnarray}
\langle n\rangle &=&\sum_{n=1}^{N}\frac{n}{N}C_{n}^{N}\psi _{n}^{2}
\label{AVN} \\
\langle nn\rangle &=&\sum_{n=1}^{N}\frac{C_{2}^{n}}{C_{2}^{N}}C_{n}^{N}\psi
_{n}^{2}  \label{AN2} \\
\langle nnn\rangle &=&\sum_{n=1}^{N}\frac{C_{3}^{n}}{C_{3}^{N}}C_{n}^{N}\psi
_{n}^{2}  \label{AN3} \\
\kappa &=&\frac{LR}{Z}\sum_{n=1}^{N}C_{n-2}^{N-2}\psi _{n-1}\psi _{n}
\label{KAP}
\end{eqnarray}%
where $\psi _{n}^{2}$ is probability for $n$ times occupancy (with $n$
electrons on the local level). Then

\begin{eqnarray}
&&\sum_{\beta }U_{\alpha \beta }\langle n_{\beta }\rangle =U(N-1)\langle
n\rangle ,  \label{FU1} \\
&&\sum_{\beta \beta ^{\prime }}U_{\alpha \beta }U_{\alpha \beta ^{\prime
}}\langle n_{\beta }n_{\beta ^{\prime }}\rangle =U^{2}\left[
(N-1)(N-2)\langle nn\rangle +(N-1)\langle n\rangle \right] ,  \label{FU2} \\
&&\sum_{\beta \beta ^{\prime }\beta ^{\prime \prime }}U_{\alpha \beta
}U_{\alpha \beta ^{\prime }}U_{\alpha \beta ^{\prime \prime }}\langle
n_{\beta }n_{\beta ^{\prime }}n_{\beta ^{\prime \prime }}\rangle =U^{3}\left[
(N-1)(N-2)(N-3)\langle nnn\rangle +3(N-1)(N-2)\langle nn\rangle
+(N-1)\langle n\rangle \right] ,  \label{FU3} \\
&&\sum_{\mathbf{k}\beta }U_{\alpha \beta }^{2}V_{\beta }(\mathbf{k})\langle
\left( 2n_{\alpha }-1\right) f_{\beta }^{\dagger }c_{\mathbf{k}\beta
}\rangle =(N-1)(ZU)^{2}(1-2\kappa )\int {\frac{d\xi }{\pi }}f(\xi )\mathrm{Im%
}\{\Delta (\xi )G_{g}(\xi )\}.  \label{FUV}
\end{eqnarray}%
Introducing the notations%
\begin{equation}
\int {\frac{d\xi }{\pi }}f(\xi )\mathrm{Im}\{\Delta (\xi )G_{g}(\xi )\}={%
\Delta G}_{g}  \label{INT}
\end{equation}%
we, for example, obtain the following formuale%
\begin{eqnarray}
\Sigma ^{(1)} &=&\left[ (N-1)(N-2)\langle nn\rangle +(N-1)\langle n\rangle
-(N-1)^{2}\langle n\rangle ^{2}\right] U^{2},  \label{SF1} \\
\Sigma ^{(2)} &=&(N-1)(N-2)(N-3)\langle nnn\rangle U^{3}+(N-1)(N-2)\langle
nn\rangle (3U^{3}+\epsilon _{f}U^{2}-2(N-1)\langle n\rangle U^{3}),  \notag
\\
&+&(N-1)^{3}\langle n\rangle ^{3}U^{3}-(N-1)^{2}\langle n\rangle
^{2}(2U^{3}+\epsilon _{f}U^{2})+(N-1)\langle n\rangle (U^{3}+\epsilon
_{f}U^{2})+(N-1)(ZU)^{2}(1-2\kappa ){\Delta G}_{g}.  \label{SF2}
\end{eqnarray}

\section{Appendix B}

In the crystal field case we assume that $N$--fold degenerate impurity
level $\epsilon _{f}$ is split by a crystal field onto $G$ sublevels
$\epsilon_{f1},...\epsilon_{f\alpha },...\epsilon_{fG}$. We assume that for each
sublevel there is still some partial degeneracy $d_{\alpha }$ so that
$\sum_{\alpha =1}^{G}d_{\alpha }=N.$ In limiting case of $SU(N)$ degeneracy,
$G=1,d_{1}=N$, and in non--degenerate case, $G=N,d_{1}=d_{\alpha }=d_{G}=1$.
We need to discuss how a number of electrons $n$ can be accommodated over
different sublevels $\epsilon_{f\alpha }$. Introducing numbers of electrons
on each sublevel, $n_{\alpha }$, we obtain $\sum_{\alpha =1}^{G}n_{\alpha
}=n.$ Note the restrictions: $0<n<$ $N,$ and $0<n_{\alpha }\leq
d_{\alpha }$. In $SU(N)$ case, $G=1,n_{1}=n$, and in non--degenerate case, $G=N$,
$n_{\alpha }$ is either 0 or 1. Total energy for the shell with $n$ electrons
depends on particular configuration $\{n_{\alpha }\}$
\begin{equation}
E_{n_{1}...n_{G}}=\sum_{\alpha =1}^{G}\epsilon _{f\alpha }n_{\alpha }+\frac{1
}{2}U(\Sigma _{\alpha }n_{\alpha })[(\Sigma _{\alpha }n_{\alpha })-1].
\label{SLAcrfLEV}
\end{equation}
Many--body wave function is also characterized by a set of numbers
$\{n_{\alpha }\}$, i.e. $|n_{1}...n_{G}\rangle .$ Energy $E_{n_{1}...n_{G}}$
remains degenerate, which can be calculated as product of how many
combinations exists to accommodate electrons in each sublevel, i.e. $%
C_{n_{1}}^{d_{1}}\times ...C_{n_{\alpha }}^{d_{\alpha }}\times
...C_{n_{G}}^{d_{\alpha }}.$ Let us further introduce probabilities $\psi
_{n_{1}...n_{G}}$ to find a shell in a given state with energy $%
E_{n_{1}...n_{G}}.$ Sum of all probabilities should be equal to $1$, i.e.%
\begin{equation}
\sum_{n_{1}=0}^{d_{1}}...\sum_{n_{\alpha }=0}^{d_{\alpha
}}...\sum_{n_{G}=0}^{d_{G}}C_{n_{1}}^{d_{1}}...C_{n_{\alpha }}^{d_{\alpha
}}...C_{n_{G}}^{d_{G}}\psi _{n_{1}...n_{G}}^{2}=1.  \label{SLAcrfSUM}
\end{equation}

There are two Green functions in Gutzwiller method: impurity Green function $%
\hat{G}_{f}(i\omega )$ and quasiparticle Green function $\hat{G}_{g}(i\omega
)=\hat{b}^{-1}\hat{G}_{f}(i\omega )\hat{b}^{-1},$ where matrix coefficients $%
\hat{b}$ represent generalized mass renormalizations parameters. All
matrices assumed to be diagonal and having diagonal elements numerated as
follows: $G_{1}(i\omega ),...G_{\alpha }(i\omega ),...G_{G}(i\omega ).$ Each
element in the Green function is represented as follows%
\begin{equation}
G_{g\alpha }(i\omega )=\frac{1}{i\omega -\lambda _{\alpha }-b_{\alpha
}^{2}\Delta _{\alpha }(i\omega )},  \label{SLAcrfGRF}
\end{equation}%
\begin{equation}
G_{f\alpha }(i\omega )=b_{\alpha }^{2}G_{g\alpha }(i\omega ).
\label{SLAcrfGRG}
\end{equation}%
and determines a mean number of electrons in each sublevel%
\begin{equation}
\bar{n}_{\alpha }=d_{\alpha }T\sum_{i\omega }G_{g\alpha }(i\omega
)e^{i\omega 0^{+}}.  \label{SLAcrfNAV}
\end{equation}%
The total mean number of electrons is thus: $\bar{n}=\sum_{\alpha =1}^{G}%
\bar{n}_{\alpha }.$ Hybridization function $\hat{\Delta}(i\omega )$ is the
matrix assumed to be diagonal and having diagonal elements numerated as
follows: $\Delta _{1}(i\omega ),...\Delta _{\alpha }(i\omega ),...\Delta
_{G}(i\omega ).$ Mass renormalizations $Z_{\alpha }=b_{\alpha }^{2}$ are
determined in each sublevel.

Diagonal elements for the self--energy are 
\begin{equation}
\Sigma _{\alpha }(i\omega )=i\omega -\epsilon _{f\alpha }-\Delta _{\alpha
}(i\omega )-G_{\alpha }^{-1}(i\omega )=i\omega (1-\frac{1}{b_{\alpha }^{2}}%
)-\epsilon _{f\alpha }-\frac{\lambda _{\alpha }}{b_{\alpha }^{2}}.
\label{SLAcrfSIG}
\end{equation}%
Here:%
\begin{eqnarray}
b_{\alpha } &=&R_{\alpha }L_{\alpha
}\sum_{n_{1}=0}^{d_{1}}...\sum_{n_{\alpha }=1}^{d_{\alpha
}}...\sum_{n_{G}=0}^{d_{G}}C_{n_{1}}^{d_{1}}...C_{n_{\alpha }-1}^{d_{\alpha
}-1}...C_{n_{G}}^{d_{G}}\psi _{n_{1}...n_{\alpha }...n_{G}}\psi
_{n_{1}...n_{\alpha }-1...n_{G}},  \label{SLAcrfEQB} \\
L_{\alpha } &=&\left( 1-\sum_{n_{1}=0}^{d_{1}}...\sum_{n_{\alpha
}=1}^{d_{\alpha }}...\sum_{n_{G}=0}^{d_{G}}C_{n_{1}}^{d_{1}}...C_{n_{\alpha
}-1}^{d_{\alpha }-1}...C_{n_{G}}^{d_{G}}\psi _{n_{1}...n_{\alpha
}...n_{G}}^{2}\right) ^{-1/2},  \label{SLAcrfEQL} \\
R_{\alpha } &=&\left( 1-\sum_{n_{1}=0}^{d_{1}}...\sum_{n_{\alpha
}=0}^{d_{\alpha
}-1}...\sum_{n_{G}=0}^{d_{G}}C_{n_{1}}^{d_{1}}...C_{n_{\alpha }}^{d_{\alpha
}-1}...C_{n_{G}}^{d_{G}}\psi _{n_{1}...n_{\alpha }...n_{G}}^{2}\right)
^{-1/2}.  \label{LSAcrfEQR}
\end{eqnarray}

The generalization of the non--linear equations (\ref{GUZ}) has the form%
\begin{eqnarray}
0 &=&\left[ E_{n_{1}...n_{G}}+\Lambda -(\Sigma _{\alpha }^{G}\lambda
_{\alpha }n_{\alpha })\right] \psi _{n_{1}...n_{G}}+  \notag \\
&&\sum_{\alpha =1}^{G}n_{\alpha }[T\Sigma _{i\omega }\Delta _{\alpha
}(i\omega )G_{g\alpha }(i\omega )]b_{\alpha }\left[ R_{\alpha }L_{\alpha
}\psi _{n_{1}...n_{\alpha }-1...n_{G}}+b_{\alpha }L_{\alpha }^{2}\psi
_{n_{1}...n_{\alpha }...n_{G}}\right] +  \notag \\
&&\sum_{\alpha =1}^{G}(d_{\alpha }-n_{\alpha })[T\Sigma _{i\omega }\Delta
_{\alpha }(i\omega )G_{g\alpha }(i\omega )]b_{\alpha }\left[ R_{\alpha
}L_{\alpha }\psi _{n_{1}...n_{\alpha }+1...n_{G}}+b_{\alpha }R_{\alpha
}^{2}\psi _{n_{1}...n_{\alpha }...n_{G}}\right] .  \label{SLAcrfSCF}
\end{eqnarray}

\end{widetext}


\begin{thebibliography}{99}
\bibitem{DMFT} For a review, see, A. Georges, G. Kotliar, W. Krauth, and
M.~Rozenberg, Rev. Mod. Phys. \textbf{68}, 13 (1996).

\bibitem{LDA} For a review, see, \textit{e.g.}, \textit{Theory of the
Inhomogeneous Electron Gas}, edited by S. Lundqvist and S. H. March (Plenum,
New York, 1983).

\bibitem{AnisimovKotliar} V. I. Anisimov, A. I. Poteryaev, M. A. Korotin, A.
O. Anokhin, and G. Kotliar, J.\ Phys.: Condens.\ Matter \textbf{35}, 7359
(1997).

\bibitem{LDA++} A. Lichtenstein and M. Katsnelson, Phys. Rev. B \textbf{57},
6884 (1998).

\bibitem{Georges} S. Biermann, F. Aryasetiawan, A. Georges, Phys. Rev. Lett. 
\textbf{90}, 086402 (2003).

\bibitem{LaTiO3} I.A. Nekrasov, K. Held, N. Blumer, A.I. Poteryaev, V.I.
Anisimov, and D. Vollhardt, Eur.\ Phys.\ J.\ B\textbf{18}, 55 (2000).

\bibitem{V2O3} K. Held, G. Keller, V. Eyert, D. Vollhardt, V.I. Anisimov,
Phys. Rev. Lett. \textbf{86}, 5345 (2001).

\bibitem{Licht} A. I. Lichtenstein, M. I. Katsnelson, G. Kotliar, Phys. Rev.
Lett. \textbf{87}, 067205 (2001).

\bibitem{Ce} K. Held, A.K. McMahan, R.T. Scalettar, Phys. Rev. Lett. \textbf{%
87}, 276404 (2001).

\bibitem{Nature} S. Savrasov, G. Kotliar, and E. Abrahams, Nature \textbf{410%
}, 793 (2001).

\bibitem{Science} Xi Dai, S. Savrasov, G. Kotliar, A. Migliori, H. Ledbetter
and E. Abrahams, Science \textbf{300}, 953 (2003).

\bibitem{reviews} K. Held \textit{et al.}, Psi--k Newsletter \#\textbf{56}
(April 2003), p. 65; A. I. Lichtenstein, M. I. Katsnelson, and G. Kotliar,
in \textit{Electron Correlations and Materials Properties}, ed. by A. Gonis,
N. Kioussis and M. Ciftan (Kluwer Academic, Plenum Publishers, 2002) p. 428.

\bibitem{AIM} P. W. Anderson, Phys. Rev. \textbf{124}, 41 (1961).

\bibitem{Jarrell} For a review, see, e.g., M. Jarrell, and J.\ E.\
Gubernatis, Physics Reports, \textbf{269}, 133 (1996).

\bibitem{jeschke} H. Jeschke and G. Kotliar, Rutgers University preprint.

\bibitem{rotors} S. Florens and A. Georges Phys. Rev. B \textbf{66}, 165111
(2002).

\bibitem{Haule:2001} K. Haule, S. Kirchner, J. Kroha, and P. W\"{o}lfle,
Phys. Rev. B \textbf{64}, 155111 (2001).

\bibitem{Vosko} S. H. Vosko, L. Wilk, and M. Nusair, Can. J. Phys. \textbf{58%
}, 1200 (1980).

\bibitem{Ceperley} D. M. Ceperley and B. J. Alder, Phys. Rev. Lett. \textbf{%
45}, 566 (1980).

\bibitem{Gutz} M. Gutzwiller, Phys. Rev. \textbf{134}, A923 (1964).

\bibitem{Ruck} G. Kotliar and A. E. Ruckenstein, Phys. Rev. Lett. \textbf{57}%
, 1362 (1986).

\bibitem{Fleszar} R. Fresard and G. Kotliar, Phys. Rev. B. \textbf{56},
12909 (1997).

\bibitem{Hasegawa} H. Hasegawa, Phys. Rev. B \textbf{56}, 1196 (1997).

\bibitem{Hubbard1} J. Hubbard, Proc. Roy. Soc. (London) A\textbf{281}, 401
(1964).

\bibitem{Florens} S. Florens, A. Georges, G. Kotliar, O. Parcollet, Phys.
Rev. B \textbf{66}, 205102 (2002).

\bibitem{Broyden} See, e.g., D. D. Johnson, Phys. Rev. B \textbf{38}, 12807
(1988).

\bibitem{Moments} W. Nolting, W. Borgiel, Phys. Rev. B \textbf{39},
6962(1989).

\bibitem{Pade} H. J.\ Vidberg and J. W.\ Serene, Journal of Low Temperature
Physics, \textbf{29}, 179 (1977).
\end{thebibliography}
\end{document}